\documentstyle[11pt]{article}
\newcommand{\heading}[1]{\vspace*{15mm}
{\Large\begin{center} {\bf{#1}} \end{center}}}
\renewcommand{\author}[3]{\vspace{5mm}
\begin{center}
{\normalsize \rm #1}\\    
{\normalsize \it #2}\\    
\vspace{0.65cm}\framebox[0truecm]{\rule[0cm]{0.cm}{0truecm}}
\vspace{0.35cm}\end{center} }
\def\lsim{~\rlap{$<$}{\lower 1.0ex\hbox{$\sim$}}}
\def\gsim{~\rlap{$>$}{\lower 1.0ex\hbox{$\sim$}}}

\newcommand{\apj}{{\em Astrophys. J.}}

\begin{document}
\heading{NONLINEAR EFFECTS DUE TO THE COUPLING\\
         OF LONG-WAVE MODES}

\author{BHUVNESH JAIN, EDMUND BERTSCHINGER}
       {Department of Physics, MIT, Cambridge, MA 02139 USA.}

\begin{abstract}{\baselineskip 0.4cm
The cosmological fluid equations are used to study the nonlinear
mode coupling of density fluctuations.
We find that for realistic cosmological spectra there is a
significant contribution to the nonlinear evolution on scales of interest
to large-scale structure from the long-wave part of the initial
spectrum. A consequence of this mode coupling is
that at high redshift, $z$, the nonlinear scale [defined by $\sigma(z)=1$]
can be significantly larger than a linear extrapolation would
indicate. For the standard CDM spectrum with a $\sigma_8=1$
normalization the mass corresponding to the nonlinear scale at
$z=20, 10, 5$ is about $100, 10, 3$ times (respectively) larger than the
linearly extrapolated value.
We also investigate the possibility of divergent contributions to the
density field from long-wave modes if the spectral index of the power
spectrum  $\ n<-1$. Using an approximate non-perturbative approach we
find that for $n>-3\,$ the
divergent contribution appears only in the phase. This can be related
to the large-scale bulk velocity, and clarifies
previous results from N-body simulations.
}
\end{abstract}

\bigskip
\section{Power Spectrum at Second Order}
Perturbative studies of the Newtonian cosmological fluid equations
have been used to study weakly nonlinear effects in the evolution
of density fluctuations (e.g., \cite{P}).
We have used second-order perturbation theory to study the enhancement
of power and changes in the time-dependence of characteristic
length scales. We assume that vorticity and pressure are
negligible. For the numerical results we have used the standard
CDM spectrum with $\Omega=1$, $\,H_0=50\,{\rm km\,s}^{-1}\,{\rm Mpc}^{-1}$
and $\sigma_8=1$.

The Fourier transform of the cosmological fluid equtions can be
written as:
\begin{equation}
{\partial\hat\delta\over\partial\tau}+\hat\theta=
-\int\!d^3k_1\,{\vec k\cdot\vec k_1\over k_1^2}\,
\hat\theta(\vec k_1,\tau)\,\hat\delta(\vec k - \vec k_1,\tau)\ ,
\label{eq:fluid1}
\end{equation}
\begin{equation}
{\partial\hat\theta\over\partial\tau}+{\dot a\over a}\,\hat\theta+
{6\over\tau^2}\,\hat\delta=-\int\!d^3k_1\,
k^2 {\vec k_1\cdot(\vec k-\vec k_1)\over2k_1^2 \vert \vec k-\vec
k_1\vert^2}\,\hat\theta(\vec k_1,\tau)\,\hat\theta(\vec k-\vec k_1,\tau)\ ,
\label{eq:fluid2}
\end{equation}
where
$\vec k$ is the comoving wavevector, $\tau$ is conformal time and
$a(\tau)\, $[$=1/(1+z)$] is the expansion factor.
$\hat \delta$ and $\hat\theta$ are, respectively, spatial Fourier
transforms of $\delta=\delta\rho/\bar\rho \,$ and
$\,\theta=\vec\nabla\cdot\vec v$.
Following the standard perturbative approach, we
expand $\hat\delta$ and $\hat\theta$ in a perturbation
series as: $\hat\delta(\vec k, \tau)=a\hat\delta_1(\vec k)+a^2
\hat\delta_2(\vec k)+...$.
Given the initial power spectrum, and assuming the initial fields to
be Gaussian random fields, the formal solutions for $\hat\delta(\vec k, \tau)$
can be used to obtain its power spectrum to second order:
\begin{equation}
P(k,\tau)=a^2(\tau) \, P_1(k) +a^4(\tau) \, P_2(k) , \label{eq:power}
\end{equation}
where
$P_1(k)=\, \langle\hat\delta_1 \hat\delta_1^\ast\rangle$
is the initial (linear) spectrum and
$P_2(k)=\langle\hat\delta_2 \hat\delta_2^\ast+2\,\hat\delta_1
\hat\delta_3^\ast\rangle$ is the second-order contribution.
$P_2(k)$ is an integral over wavevectors
involving products of $P_1(k_1)$ evaluated at different $k_1$.
We have calculated $P(k,\tau)$ at different values of $a$
for the standard CDM spectrum
and estimated the relative contributions from different parts of
the initial spectrum.

We find that the second-order contribution provides a significant
enhancement of power in the weakly nonlinear regime. It is in
excellent agreement with results from a high resolution N-body
simulation for $z\gsim 2$, but shows a larger enhancement for smaller
$z$. Due to the effect of long-wave mode coupling, the nonlinear
evolution responds to the shallower slope of the spectrum at long
waves. This slows the growth in time of characteristic nonlinear scales,
i.e., causes them to be larger at early times.
We find that the mass
corresponding to the scale at which the r.m.s. fluctuation $\sigma(z)=1$
at $z=20, 10, 5$ is about $100, 10, 3$ times (respectively) larger than
indicated by the standard practice of
extrapolating the nonlinear scale today using the linear
spectrum. This result has significant implications for
nonlinear structures at high $z$.

\section{Coupling of Long-Wave Modes}
An alternative to the perturbative approach is to approximate
the nonlinear terms on the right-hand side of the fluid
equations shown above so as to include the dominant contribution
from the long-wave modes. In the limit that the wavevector
$k_1\to 0$ one finds that the mean square of the nonlinear terms
takes the form
$ P_{1}(k)\!\int\!dk_1\,P_{1}(k_1).$
This integral is divergent as $k_1\to0$ if $P_{1}(k_1)\propto k_1^n$
with $n<-1$, even while the density variance is finite for $n>-3$.
Indeed some studies of N-body simulations indicate
that the evolution of $n<-1$ spectra is qualitatively different
from that of $n>-1$ spectra, and is strongly influenced by the
size of the box (the small-$k$ cutoff) \cite{RG}, \cite{G}.

The nonlinear terms described above can be approximated by
making a Taylor series expansion in ($k_1/k$) so as to include
the dominant contribution from small $k_1$. On further approximating
the small-$k_1$ density and velocity fields by the linear fields,
the equations reduce to linear partial differential equations.
We have separated the equations for the phase and amplitude
of the density field retaining the first order Taylor series term, and
have obtained a closed form solution for the phase.

 The divergent nonlinear terms drop out of the
equation for the density amplitude implying that the
power spectrum does not diverge for $-3<n<-1$ in this approximation.
 The solution for the phase is divergent for
$n<-1$. This divergence is related to the divergence of the
linear mean square bulk velocity, and corresponds to translational
motion of the fluid past the fixed coordinate frame.
It agrees exactly with the deviation of the phases from their
initial values found in N-body simulations.
However this deviation does not seem to indicate dynamically interesting
nonlinear evolution as it does not affect the amplitude.
This conclusion appears to differ from the results of some $n=-2$
simulations. We are currently investigating the
source of disagreement and also analyzing the long-wave limit of
perturbative contributions to the power spectrum to verify the
above results.\\
{\bf Acknowledgements.}
We are grateful to Alan Guth for several useful discussions.

\vfill
\end{document}